\documentclass[twocolumn]{aastex631}

\DeclareRobustCommand{\uppartial}{\text{\rotatebox[origin=t]{15}{\scalebox{0.98}[1]{$\partial$}}}\hspace{-0.5pt}}

\newcommand{\msunh}{\>h^{-1}\rm M_\odot}
\newcommand{\mpch}{\>h^{-1}{\rm {Mpc}}}
\newcommand{\rmm}{{\rm m}}

\usepackage[T1]{fontenc}
\usepackage{graphicx}	
\usepackage{amsmath}	
\usepackage{amssymb}	
\usepackage{bm}

\begin{document}

\title{(DarkAI) Mapping the large-scale density field of dark matter using artificial intelligence}

\author{Zitong Wang}
\affiliation{School of Aerospace Science and Technology, Xidian University, Xi'an 710126, People's Republic of China}

\author{Feng Shi}
\affiliation{School of Aerospace Science and Technology, Xidian University, Xi'an 710126, People's Republic of China}

\author{Xiaohu Yang}
\affiliation{Department
of Astronomy, School of Physics and Astronomy, and Shanghai Key Laboratory for Particle Physics and Cosmology,\\ 
~~Shanghai Jiao Tong University, Shanghai 200240, People's Republic of China }
\affiliation{Tsung-Dao Lee Institute and Key Laboratory for
    Particle Physics, Astrophysics and Cosmology, Ministry of Education, \\
    Shanghai Jiao Tong University, Shanghai {\rm 200240}, People's Republic of China}

\author{Qingyang Li}
\affiliation{Department
of Astronomy, School of Physics and Astronomy, and Shanghai Key Laboratory for Particle Physics and Cosmology,\\ 
~~Shanghai Jiao Tong University, Shanghai 200240, People's Republic of China }
\affiliation{Institute for Astronomy, University of Edinburgh, Royal Observatory, Edinburgh EH9 3HJ, United Kingdom}

\author{Yanming Liu}
\affiliation{School of Aerospace Science and Technology, Xidian University, Xi'an 710126, People's Republic of China}

\author{Xiaoping Li}
\affiliation{School of Aerospace Science and Technology, Xidian University, Xi'an 710126, People's Republic of China}



\begin{abstract}

Herein, we present a deep-learning technique for reconstructing the dark-matter density field from the redshift-space distribution of dark-matter halos. We built a UNet-architecture neural network and trained it using the COmoving Lagrangian Acceleration fast simulation, which is an approximation of the N-body simulation with $512^3$ particles in a box size of $500 \mpch$. Further, we tested the resulting UNet model not only with training-like test samples but also with standard N-body simulations, such as the Jiutian simulation with $6144^3$ particles in a box size of $1000 \mpch$ and the ELUCID simulation, which has a different cosmology. The real-space dark-matter density fields in the three simulations can be reconstructed reliably with only a small reduction of the cross-correlation power spectrum at 1\% and 10\% levels at $k=0.1$ and $0.3~h\mathrm{Mpc^{-1}}$, respectively. The reconstruction clearly helps to correct for redshift-space distortions and is unaffected by the different cosmologies between the training (Planck2018) and test samples (WMAP5). Furthermore, we tested the application of the UNet-reconstructed density field to obtain the velocity \& tidal field and found that this approach provides better results compared to the traditional approach based on the linear bias model, showing a 12.2\% improvement in the correlation slope and a 21.1\% reduction in the scatter between the predicted and true velocities. Thus, our method is highly efficient and has excellent extrapolation reliability beyond the training set. This provides an ideal solution for determining the three-dimensional underlying density field from the plentiful galaxy survey data.

\end{abstract}



\section{Introduction} \label{sec:intro}

Mapping the distributions of galaxies offers a key observational probe for the large-scale mass distribution in the universe, thereby constraining cosmological models \citep{1994MNRAS.267..927F,2001Natur.410..169P,2003MNRAS.346...78H,2008ApJ...676..248Y,2005ApJ...631...41T,2018ApJ...861..137S}, providing information on galaxy formation \citep{1998ApJ...494....1J,2000MNRAS.318.1144P,2003MNRAS.339.1057Y,2012ApJ...752...41Y} and facilitating the exploration of the universe's evolution \citep{2014ApJ...794...94W, 2016ApJ...831..164W,2017ApJ...841...55T}. Therefore, one of the primary goals of the redshift surveys of galaxies, such as the 2-degree Field Galaxy Redshift Survey \citep{2001MNRAS.328.1039C} and Sloan Digital Sky Survey (SDSS) \citep{2000AJ....120.1579Y} is to provide a database for studying the three-dimensional distribution of galaxies as accurately as possible. In the next decade, future galaxy surveys based on instruments such as Dark Energy Spectroscopic Instrument \citep{2016arXiv161100036D,2016arXiv161100037D}, Large Synoptic Survey Telescope \citep{2019ApJ...873..111I}, EUCLID \citep{2011arXiv1110.3193L}, Wide Field Infrared Survey Telescope \citep{2019arXiv190205569A}, and Chinese Space Station Telescope \citep{2011SSPMA..41.1441Z,2019ApJ...883..203G} will map out an unprecedented large volume of the universe with extraordinary precision. Thus, it is vital to have an optimum approach that can accurately and efficiently determine the three-dimensional underlying density field from the rich galaxy survey data.

However, a major issue with this endeavor is that galaxies are biased tracers of mass distribution, and one has to understand the connection between galaxies and dark matter before using the galaxy distribution in space to analyze the mass distribution in the universe. In the past two decades, a tremendous amount of effort has been put into establishing the relationship between galaxies and dark-matter halos, as parameterized by the conditional luminosity function or the halo occupation distribution \citep{1998ApJ...494....1J,2000MNRAS.318.1144P,2003MNRAS.339.1057Y,2003MNRAS.340..771V,2007MNRAS.376..841V,2005ApJ...633..791Z,2005ApJ...631...41T,2006MNRAS.372..758M,2008ApJ...682..937B,2009MNRAS.392..801M,2009MNRAS.394..929C,2011MNRAS.414.1405N,2011ApJ...736..134A,2012ApJ...744..159L}. An empirical way to establish the galaxy--halo connection is to use galaxy groups, which are sets of galaxies that reside in the same dark-matter halo. \citep{2005MNRAS.356.1293Y,2007ApJ...671..153Y,2021ApJ...909..143Y} developed a halo-based group identifier that is optimized for classifying galaxies that reside in the same dark-matter halos, making it ideal for studying the relationship between galaxies and dark-matter halos. The well-understood galaxy-halo relationship allows us to reconstruct the underlying cosmic density field using the dark halos represented by galaxy systems, thereby enabling the detailed study of the relationships among galaxies, dark halos, and large-scale structures (LSSs).

Reconstructing the cosmic density field from galaxy distribution has been conducted earlier based on several redshift surveys \citep{1995MNRAS.272..885F,1995ApJ...449..446Z,1999AJ....118.1146S,2002MNRAS.333..739M,2004MNRAS.352..939E,2009MNRAS.394..398W,2013ApJ...772...63W}. In these investigations, the distribution of galaxies is generally heavily smoothed and normalized to represent the cosmic density field on large scales. The Wiener reconstruction method, employed in \citep{1995MNRAS.272..885F} and \citep{1995ApJ...449..446Z}, can generate a reconstructed density field with the minimum mean square error by assuming that the mass density at a given point is a linear combination of the observed galaxy density-field values at various points. The reconstruction method employed in \citep{2009MNRAS.394..398W,2012MNRAS.420.1809W,2013ApJ...772...63W} is based on dark-matter halos represented by galaxy groups \citep{2005MNRAS.356.1293Y, 2007ApJ...671..153Y, 2012ApJ...752...41Y} and considers that the amplitude of the halo density field is linearly biased to that of the underlying density field. Notably, the density reconstruction method that we adopt for the present-day, nonlinear density field is not used for the baryon acoustic oscillations (BAO) (e.g., \citep{2007ApJ...664..675E}), which requires the initial, linear density field. Furthermore, in the linear theory of gravitational instability, the gravity and velocity vectors are parallel and related to each other by a proportionality constant that depends only on the mean mass density of the universe, $\Omega_\rmm$. If one can reconstruct the density field from the galaxy distribution, it is possible to determine the gravity field from the distribution and then infer the linear velocity field \citep{2009MNRAS.394..398W,2013ApJ...772...63W}.

\begin{figure*}
    \centering
	\includegraphics[width=2.0\columnwidth]{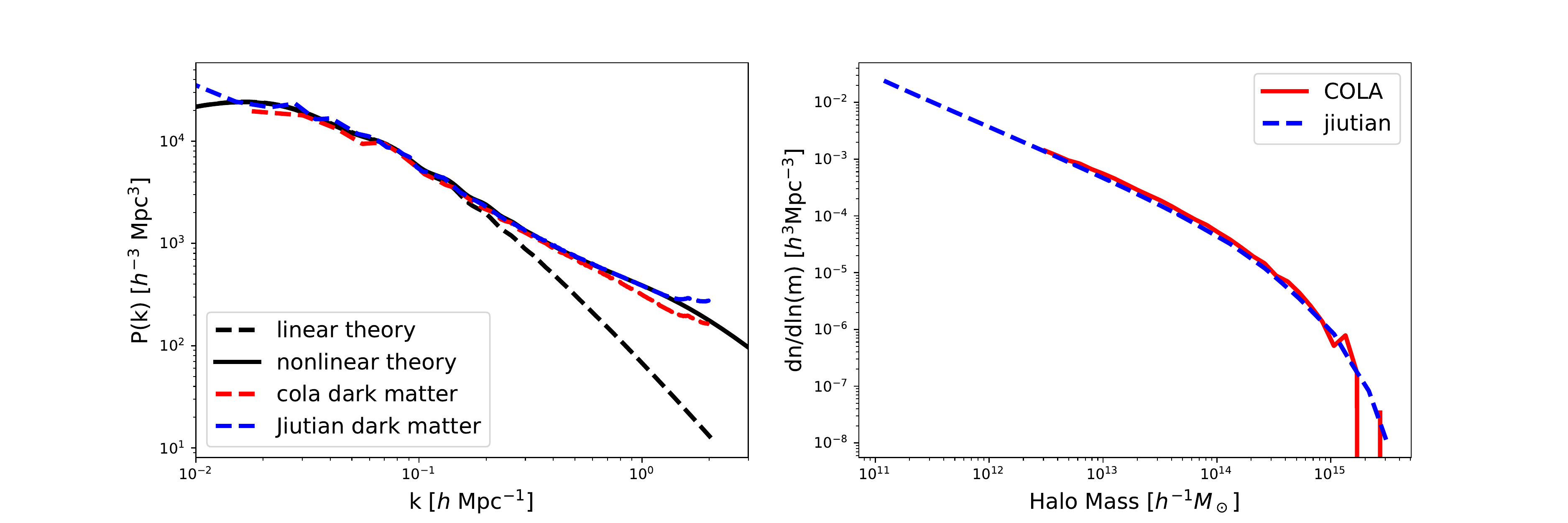}
    \caption{(Left panel) Power spectra of the dark-matter density field. The result of the COmoving Lagrangian Acceleration (COLA) simulation with 500 $\mpch$ box size and $512^3$ particles is shown by the red dashed line, and that of the Jiutian simulation with 1000 $\mpch$ box size and $6144^3$ particles is represented by the blue dashed line. The nonlinear and linear power spectra are represented by black solid and dashed lines, respectively. (Right panel) Halo mass functions obtained for the COLA (red solid) and Jiutian (blue dashed) simulations.}
    \label{fig:cola_jiutian_pkhm}
\end{figure*}

However, reconstructing the density and velocity fields from galaxy groups or halos is fraught with complications. First, as previously stated, some constraints for relating the fluctuations in the halo distribution to those in the mass distributions must be adopted, assuming that the two are related by a linear bias parameter $b$. However, this model is motivated more by simplicity than physical principles because the bias parameter $b$ is dependent on the scale (e.g., \citep{2007JCAP...10..007C,2018A&A...613A..15S}). The simple proportionality between the gravity and velocity fields in linear theory is valid only when the density fluctuations are small. Once the clustering of dark matter becomes nonlinear, this one-to-one correspondence ceases due to shell crossing. Another problem emerges from the redshift-space distortions (RSDs): the spatial distribution of galaxies observed in the redshift space is distorted with respect to the real-space distribution \citep{1977ApJ...212L...3S,1983ApJ...267..465D,1987MNRAS.227....1K,1992ApJ...385L...5H} the redshifts of galaxies are not precise measures of distances because of the peculiar motions of galaxies. In terms of large scales, coherent flows induced by the gravitational action of LSS improve the structure along the line-of-sight (the Kaiser effect) \citep{1987MNRAS.227....1K}. Meanwhile, in terms of small scales, the virialized motion of galaxies within dark-matter halos blurs the structure along the line-of-sight (the Finger-of-God (FOG) effect) \citep{1972MNRAS.156P...1J,1978IAUS...79...31T}. The approaches adopted earlier to deal with RSDs during density-field reconstruction were limited because the large-scale Kaiser effect and the small-scale FOG effect are intertwined, rendering the modeling of the bias parameters more complicated with their forms being unknown a priori \citep{2016ApJ...833..287L, 2016ApJ...833..241S}. Numerous methods for distortion correction are available to reconstruct the real-space density and velocity fields. The studies that employ either linear theory or the Zeldovich approximation are limited in their ability to reconstruct high-density regions. Thus, to fully exploit galaxy redshift surveys for mapping the LSS of the universe, a change in strategy is required.

In recent years, deep-learning methods have been employed for studying cosmology and LSS; for example, estimating cosmological parameters from the dark-matter distribution \citep{2017arXiv171102033R,2020SCPMA..6310412P}, predicting cosmological structure formation \citep{2018MNRAS.479.3405L,2019PNAS..11613825H}, populating the halo position and mass field from the nonlinear matter field \citep{2018JCAP...10..028M,2019MNRAS.490..331L, 2019MNRAS.482.2861B,2019PhRvD.100d3515K}, classifying the LSS of the universe \citep{2019MNRAS.484.5771A}, simulating the underlying luminosity function of line intensity maps \citep{2019arXiv190510376P}, analyzing the relation between galaxy distribution in hydrodynamic simulations and its underlying dark-matter distribution \citep{2019arXiv190205965Z,2019MNRAS.487L..24T}, reconstructing the BAO signal \citep{2021MNRAS.501.1499M}, and reconstructing the nonlinear velocity field \citep{2021ApJ...913....2W,2023arXiv230104586W}. Several studies have employed these models to understand the physics of LSS because of their ability to learn complex functions. However, training a neural network (NN) necessitates a concerted effort of accurate and fast predictions of the universe's structure. An N-body simulation is an effective method for predicting the formation of the universe's structure; however, it presents a significant computational challenge while evolving the dynamics of billions of particles involved in the simulation.

Herein, we present a deep-learning technique for predicting the real-space density field of dark matter from the redshift-space density field of dark-matter halos. We perform the training using a fast N-body simulation generated by the COmoving Lagrangian Acceleration (COLA) method \citep{2013JCAP...06..036T}. Unlike traditional N-body methods, the COLA method can directly trade accuracy at small scales for computational speed while retaining accuracy at large scales. This is particularly useful for quickly and affordably generating large ensembles of accurate mock halo catalogs required to study the reconstruction of the dark-matter density field. We demonstrate that the NN, which is trained using a limited set of COLA samples, can achieve accurate extrapolation beyond its training data.

This paper is structured as follows: In Section~\ref{sec:data}, we present the dataset used to generate the training and testing samples used herein. In Section~\ref{sec:network}, we present a method for reconstructing the dark-matter density field using artificial intelligence (AI). The results of the accuracy evaluation of the reconstructed density field are presented in Section~\ref{sec:val_cola} and Section~\ref{sec:val_jiutian}. Further, we discuss the application of the AI-reconstructed density field for velocity \& tidal field reconstruction in Section~\ref{sec:val_vectid}. Finally, we summarize our main findings in Section~\ref{sec:con}.

\begin{figure*}
	\hspace*{1.8cm}\includegraphics[width=1.5\columnwidth]{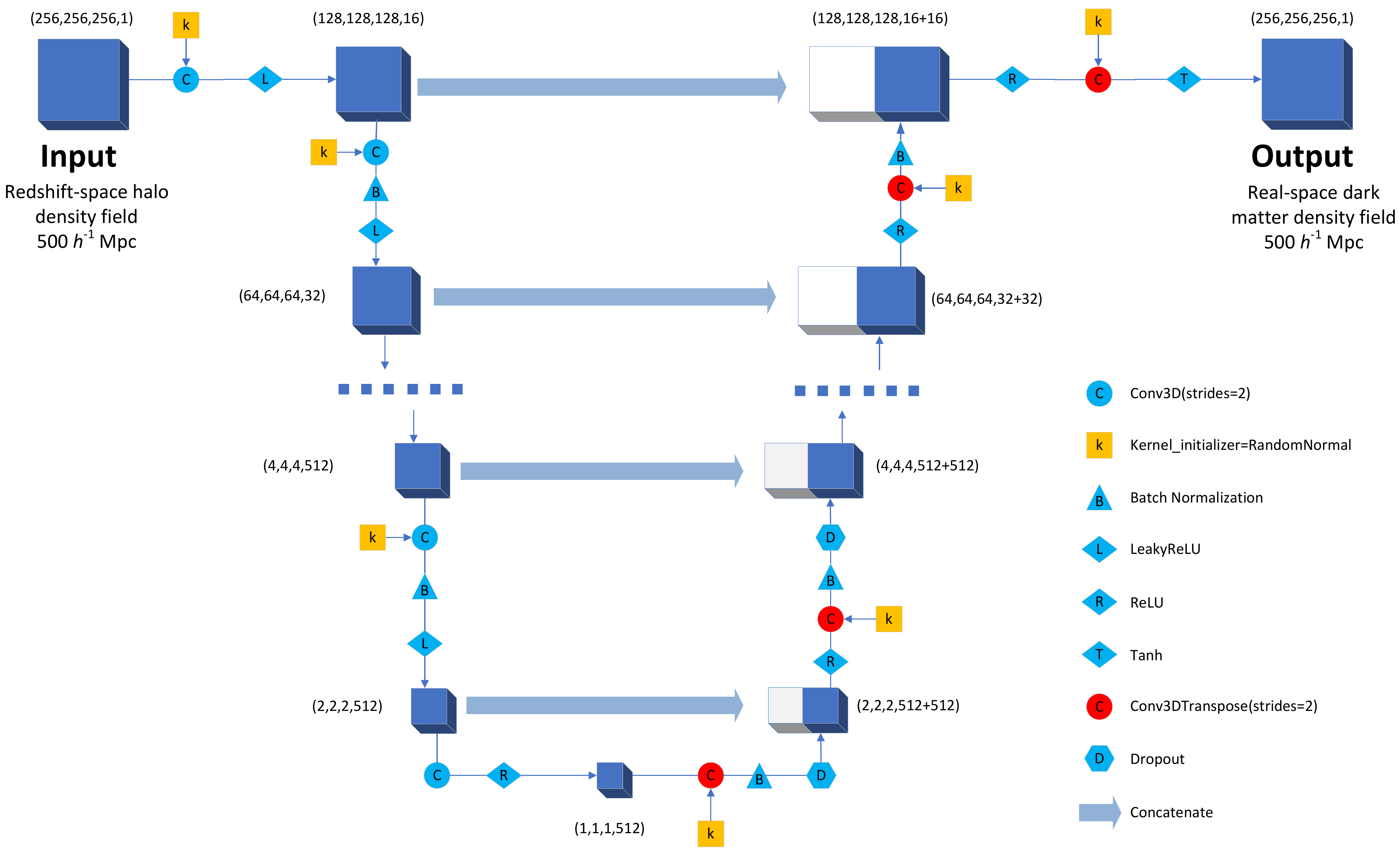}
    \caption{Three-dimensional (3D) UNet architecture. Each cube represents a 3D space-with-feature map with the grid size and the number of feature channels indicated. The input and output cubes, each with $256^3$ voxels and 500 $\mpch$ in size, correspond to the redshift-space halo density field and the real-space dark-matter density field, respectively. The encoder employs seven convolutional layers to downsample the input cube to a bottleneck, while the decoder part reads the bottleneck output and employs seven transposed convolutional layers to upsample to the output cube. Each layer in the encoder employs Convolution-BatchNorm-LeakyRelu blocks, while those in the decoder employ Convolution-BatchNorm-Dropout-ReLU blocks with a dropout rate of 50\%. For clarity, only the blocks for the first and last layers of the encoder and decoder are shown. Calculations are represented by different symbols, as shown in the lower-right corner.}
    \label{fig:network}

    \hspace*{0.8cm}\includegraphics[trim=4cm 0cm 0cm 0cm,clip=True,width=2.1\columnwidth]{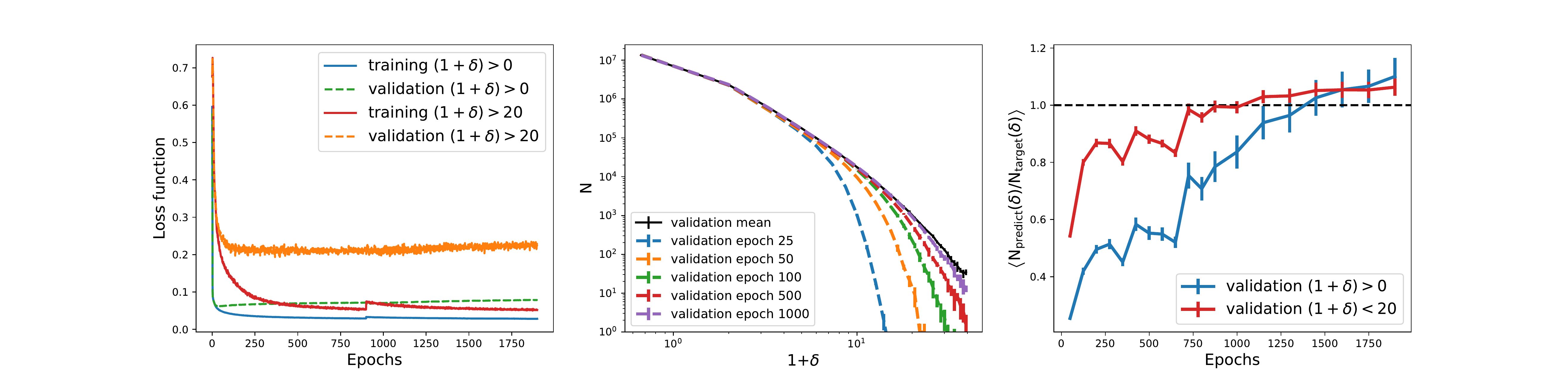}
    \caption{(Left panel) Evolution of the loss function. The blue solid and green dashed lines represent the loss functions in the training and validation samples, respectively, which are derived from all the voxels with $(1+\delta) > 0$. The loss functions derived from the voxels with $(1+\delta) > 20$ are provided for the training (red solid) and validation (orange dashed) samples. (Middle panel) Comparison of the density distributions for the validation samples. The y-axis represents the number of voxels in each $1+\delta$ bin. The result averaged from the five validation samples is indicated by a black solid line. Different colored lines represent the predicted fields at epochs 25, 50, 100, 500, and 1000. The error bars represent the variance among the five validation samples. (Right panel) Averaged ratios of the density distribution between the prediction and target as a function of epochs. The blue and red lines represent validation-set results from the voxels with $(1+\delta) > 0$ and $(1+\delta) < 20$, respectively.}
    \label{fig:trainloss}

    \hspace*{-0.55cm}\includegraphics[trim=5cm 0cm 3.5cm 0cm,clip=True,width=2.2\columnwidth]{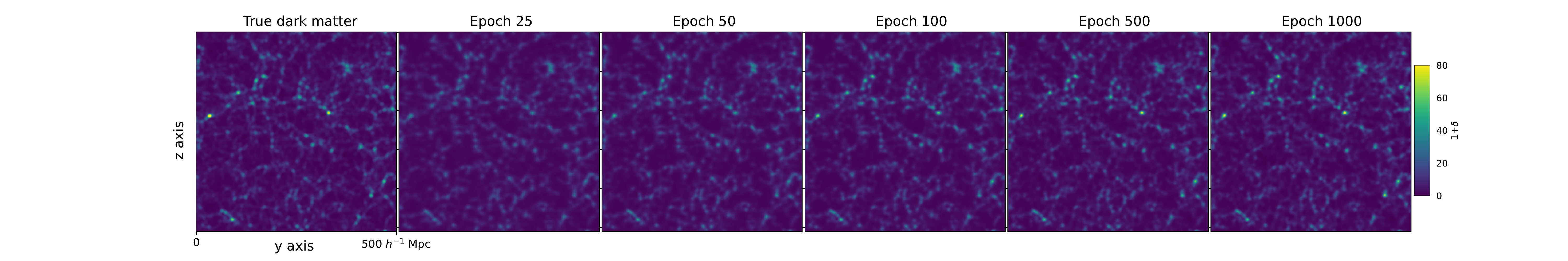}
    \caption{Comparison of the projected density-field distributions in a slice of $500 \times 500 \times 9.76 \mpch$ for a validation sample. The first panel shows the true dark-matter density fields derived from the original simulation. The other five panels represent the predictions at epochs 25, 50, 100, 500, and 1000.}
    \label{fig:trainslices}
\end{figure*}
%
\section{DATASETS}
\label{sec:data}
The training and testing samples were generated using the {\tt cola\_halo}\footnote{Downloaded from \url{https://github.com/junkoda/cola_halo}} code \citep{2016MNRAS.459.2118K}. The code generates random Gaussian initial conditions based on the second-order Lagrangian Perturbation Theory (2LPT) \citep{2006MNRAS.373..369C}, realizes N-body particle evolution by the COLA \citep{2013JCAP...06..036T} fast simulation method, and identifies dark-matter halos using the friends-of-friends algorithm \citep{1985ApJ...292..371D}. By combining 2LPT and the N-body code, which are used to solve for large- and small-scale dynamics, respectively, a COLA-based simulation run that is orders of magnitude faster than a standard detailed N-body simulation can be achieved.

We generated a set of 30 simulations with independent initial conditions. The cosmological parameters were set to be consistent with the Planck2018 results \citep{2020A&A...641A...6P}: $\Omega_\rmm = 0.3111$, $\Omega_\Lambda = 0.6889$, $h=0.6766$, and $\sigma_8=0.812$. Each simulation evolved with $512^3$ dark-matter particles in a periodic box of $500 \mpch$. The mass of each particle was $8.62 \times 10^{10} \msunh$. Dark-matter halos were identified using the standard friends-of-friends (FOF) algorithm with a linking length that is 0.2 times the mean interparticle separation. The halo mass was defined as the sum of the masses of all particles in the halo, and halos with <32 particles were removed. Therefore, the minimum mass of the COLA halos was $M_\mathrm{min} \approx 10^{12.44}\msunh$. The particle and halo distributions at redshift $z=0$ were used, and the redshift-space distributions of the halo samples were constructed by assuming the z-axis as the line-of-sight. According to the SDSS group catalog \citep{2007ApJ...671..153Y,2012MNRAS.420.1809W}, mass assignment based on the characteristic luminosity ranking is complete to $z \sim 0.14$ for groups with halo masses of $M \gtrsim 10^{12.5}\msunh$. In our COLA simulations, we used halos with masses of $M \geqslant 10^{12.5}\msunh$ as primary samples and used two further samples with halo masses of $M \geqslant 10^{13.0}\msunh$ and $10^{13.5}\msunh$ for comparisons.

We implemented the cloud-in-cell \citep{1981csup.book.....H} scheme to calculate the overdensity, $\delta$, on a grid of $256^3$ voxels for the dark matter and halo distributions. Further, we applied a top-hat smoothing kernel to the overdensity fields with a smoothing scale of $R_{\rm s}= 5\mpch$, which is an optimal value in our model based on tests. To increase the contrast of the voxel values, we rescaled the overdensity values to lie in the interval $[-1, 1]$ by the following nonlinear transformation \citep{2018ComAC...5....4R}: $s(x) = 2x/(x + a) - 1$, where $a = 5$. Each dark-matter halo is assigned a mass weight to determine the halo density field. We divided the dataset into 15 (training), 5 (validation), and 10 (test) samples for each of the halo samples.

To verify the generality of our results, we used a high-resolution N-body simulation, Jiutian, as a test sample. This simulation was run at the High-Performance Computing Center at Kunshan using L-GADGET, a memory-optimized version of the GADGET2 code \citep{2005MNRAS.364.1105S}. The Jiutian simulation describes the distribution of $6144^3$ dark-matter particles in a periodic box of $1000 \mpch$. The cosmological parameters used in this simulation were also consistent with the Planck2018 results. Dark-matter halos were detected using the FOF algorithm with the same linking length as the COLA simulation (Han et al. 2023, in preparation). Figure~\ref{fig:cola_jiutian_pkhm} shows the dark-matter power spectra and halo mass function obtained for the COLA and Jiutian simulations at redshift $z=0$. Both simulations are consistent with each other in terms of the power spectra at the median scale and halo mass distributions above $10^{12.5}\msunh$. However, due to box size and resolution limitations, the COLA results are expected to be distinct from the Jiutian results in terms of large and small scales. Hence, the objective of this paper is to test the applicability of the model that is trained using low-resolution simulations before applying to the high-resolution simulations. To calculate the overdensity field in the Jiutian case, we created a grid of $512^3$ voxels and divided it into eight small grids of $256^3$ voxels each. Additionally, similar to the COLA simulation, we apply the smoothing and rescaling procedure along with the halo mass weight strategy. To ensure consistency with the main halo samples used in the COLA simulation, we only consider the halos with mass $M \geqslant 10^{12.5}\msunh$ in the Jiutian simulation.

Additionally, we used the N-body simulation ELUCID \citep{2016ApJ...831..164W}, which has a different cosmology, as a test sample to study the influence of cosmology on our reconstructions. This simulation, which was run with L-GADGET at the Center for High-Performance Computing, Shanghai Jiao Tong University, achieved the evolution of the distribution of $3072^3$ dark-matter particles in a periodic box of $500 \mpch$ on a side. The cosmological parameters used in the simulation were consistent with the WMAP5 results: $\Omega_\rmm = 0.258$, $\Omega_\Lambda = 0.742$, $h=0.72$, and $\sigma_8=0.80$. The overdensity field of dark-matter particles and halos was obtained in the same way as for the COLA main samples, where the halo mass was $M \geqslant 10^{12.5}\msunh$.

\section{NEURAL NETWORK}
\label{sec:network}

We adapt our network architecture following the UNet style \citep{2015arXiv150504597R} and modify its design following the Pix2Pix generator \citep{2016arXiv161107004I}. Figure~\ref{fig:network} shows the main components of the architecture. Each cube represents a 3D space-with-feature map, with the grid size and number of feature channels indicated. The input and output cubes, each with $256^3$ voxels and 500 $\mpch$ in size, correspond to the redshift-space halo density field and the real-space dark-matter density field. The encoder section comprises seven convolutional layers that downsample the input cube, reducing the spatial size while increasing the feature channel. Skip connections are used to connect layers in the encoder to corresponding layers in the decoder, allowing spatial information to be preserved during downsampling. The decoder part then upsamples to the required output cube size using seven transposed convolutional layers. The detailed operations are represented by various symbols, as shown in Figure~\ref{fig:network}.

Furthermore, we tested the Pix2Pix full architectures with an additional discriminator that has been trained to detect the generator's ``fakes'' as well as possible. The advantage of this architecture may be a better ability to forecast small-scale structure \citep{2016arXiv161107004I}. However, in practice, Pix2Pix as a generative adversarial network suffers from mode collapse, a common issue that complicates and increases the likelihood of training failure \citep{2020arXiv200209124F}. Herein, we will show how an architecture based solely on the Pix2Pix generator from Figure~\ref{fig:network} can still produce a reliable result on a small scale.

To train the networks, we used the mean absolute error as the loss function,
\begin{align}\label{L1_equation}
    \mathcal{L}(\delta_\mathrm{p},\delta_\mathrm{t}) = \mathbb{E}[|\delta_\mathrm{p}-\delta_\mathrm{t}|],
\end{align}
where $\delta_\mathrm{p}$ refers to the predicted density field and $\delta_\mathrm{t}$ refers to the simulation target. We trained the networks thoroughly, with weights drawn from a Gaussian distribution with mean and standard deviation values of 0 and 0.02, respectively. The network uses 15 training samples in each epoch, followed by five validation samples that are used only to evaluate the models during training.

In the left panel of Figure~\ref{fig:trainloss}, we show the evolution of the loss function for 1900 epochs. The blue solid and green dashed lines represent the loss functions from the training and validation samples, respectively, which are derived from all the voxels with $(1+\delta) > 0$. Furthermore, the loss functions derived from voxels with $(1+\delta) > 20$ are provided for training (red solid) and validation (orange dashed). It demonstrates that the all-voxel loss function of the validation set converges after about 50 epochs. However, one can see that the loss function of the validation set continues to decrease after epoch 50 and converges after about 250 epochs when $(1+\delta) > 20$. This demonstrates that the all-voxel loss function, whose values are dominated by a large number of small densities, makes evaluating the training performance of the high-density regions difficult. For example, only 0.02\% of total voxels have $(1+\delta) > 20$. As an illustration, Figure~\ref{fig:trainslices} shows the comparisons of projected density-field distributions in a slice of $500 \times 500 \times 9.76 \mpch$ for one validation sample. The first panel shows the true dark-matter density fields obtained from the original simulation. The remaining five panels represent predictions at epochs 25, 50, 100, 500, and 1000. Although it is similar to the target on a large scale, the prediction at epoch 25 cannot present the high-density region, such as the cluster structure. The high-density structure becomes more visible at later epochs, and the prediction produced at epoch 1000 outperforms that produced at earlier epochs. As a result, the networks can still learn effectively even at the convergence stage of the all-voxel loss function.

\vspace{-0.01cm}

In this case, we compare the density distribution of the validation samples between the targeted and predicted fields at epochs 25, 50, 100, 500, and 1000, as shown in the middle panel of Figure~\ref{fig:trainloss}. Evidently, from epoch 25 to 1000, there are smaller disparities at the $(1+\delta) > 20$ between the predictions and targets, indicating that high-density voxel recovery has significantly improved. However, as training processes, the predictions appear to have an unreasonably high density. In the right panel of Figure~\ref{fig:trainloss}, we show the averaged ratios of the density distribution between predictions and targets as a function of epochs. The blue and red lines represent validation-set results from voxels with $(1+\delta) > 0$ and $(1+\delta) < 20$, respectively. Although the all-voxel ratios reach unity at around epoch 1300, a significant fraction of voxels with $(1+\delta) < 20$ have ratios greater than unity. We confirmed that this would result in excessive clustering in the density field using the model saved at this stage. Therefore, considering the voxels with $(1+\delta) < 20$, we choose the model as our best model based on the validation-set ratios that were closest to unity between epochs 600 and 1000.

Following training, we validated the resulting model using test samples from the COLA fast, Jiutian, and ELUCID N-body simulations, as described in Section~\ref{sec:data}.
\begin{figure*}
    \hspace*{-0.5cm}\includegraphics[trim=4cm 0cm 2cm 0cm,clip=True,width=2.2\columnwidth]{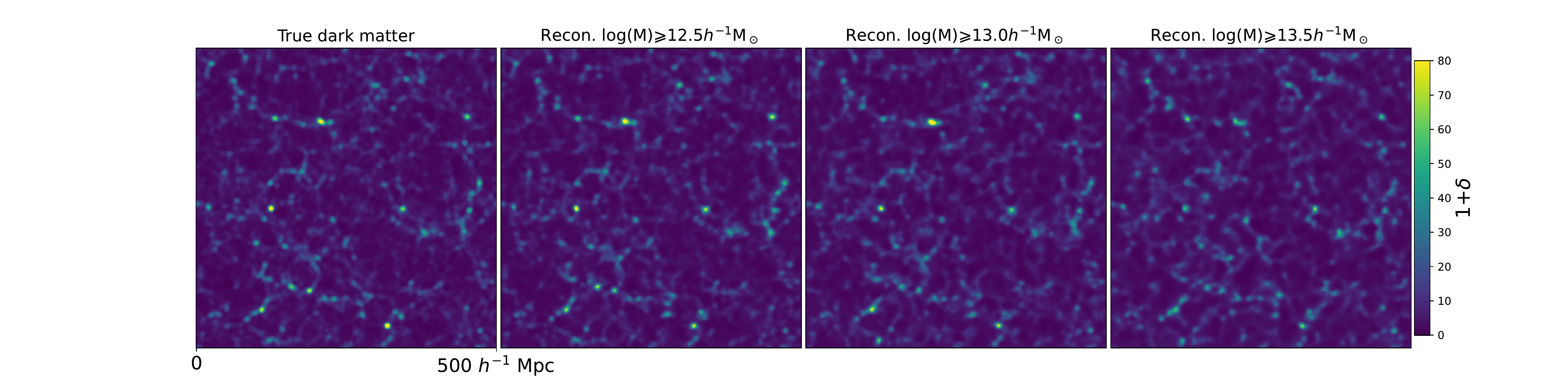}
    \caption{COLA test: Comparison of the projected density-field distributions in a slice of $500 \times 500 \times 9.76 \mpch$. The first panel shows the true dark-matter density field. The other three panels show the UNet-reconstructed dark-matter density field obtained using halos with masses above the mass thresholds of $12.5$, $13.0$, and $13.5$ (log($\msunh$)), respectively.}
    \label{fig:slices_cola}

    \includegraphics[width=2.05\columnwidth]{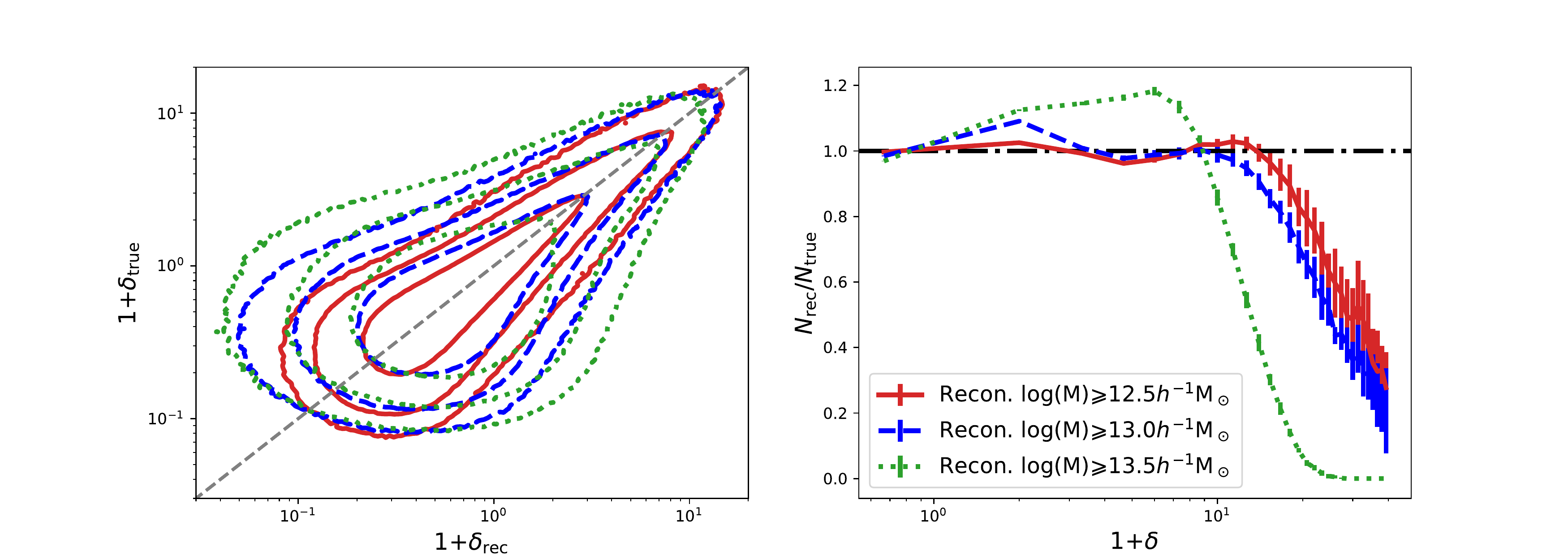}
    \caption{COLA test: (Left panel) Density--density relationship exists between the reconstructed density field $1+\delta_\mathrm{rec}$ and the true field $1+\delta_\mathrm{true}$. Different colored curves correspond to different halo mass thresholds, as indicated. The three contours in each threshold case cover 67\%, 95\%, and 99\% of the grid cells. The dashed lines indicate the perfect relationship. (Right panel) Density distribution ratios of the reconstruction and truth. The error bars represent $\pm 1\sigma$ variance among the 10 test samples.}
    \label{fig:den_corr}
\end{figure*}

\section{Validation with COLA test samples}
\label{sec:val_cola}

In this section, we present the accuracy and reliability test results for our resulting model obtained using the 10 test samples from the COLA simulations.

\subsection{Pixel-to-pixel comparison}
\label{sec:cola_pixcom}

We begin with a visual comparison based on the projected density-field distribution in a slice of $500 \times 500 \times 9.76 \mpch$, as shown in Figure~\ref{fig:slices_cola}. The true dark-matter density field is shown in the first panel. From left to right, the other three panels show the UNet-reconstructed dark-matter density field obtained using halos at the corresponding mass thresholds of $12.5$, $13.0$, and $13.5$ (log($\msunh$)), respectively. When compared with the true field, the reconstructed density fields exhibit recognizable and consistent large-scale structures (such as clusters, filaments, and voids), indicating that the reconstructions are successful in general. Further, these results show that the reconstructions are generally resistant to the boundary effect because the structures are effectively recovered at the border of the box. However, high-density voxels in the reconstructions, such as those surrounding the clusters, can be shown to differ from the target, which is more noticeable when the mass threshold is 13.0 or 13.5.

The left panel of Figure~\ref{fig:den_corr} shows the density--density relationship between the reconstructed density field $1+\delta_\mathrm{rec}$ and the true field $1+\delta_\mathrm{true}$. Different colored curves correspond to different halo mass thresholds, as indicated. The three contours in each threshold case cover 67\%, 95\%, and 99\% of the grid cells. The dashed lines indicate the perfect relationship. Although there is no significant bias in any of the reconstructions, there is an increasing scatter from mass thresholds of 12.5 to 13.0 \& 13.5. The right panel of Figure~\ref{fig:den_corr} shows the ratios of density distribution between the reconstruction and the truth. The error bars represent the $\pm 1\sigma$ variance among the 10 test samples. The ratios at $1+\delta<15$, which covers 99.94\% of the grids, are quite close to unity within 5\% for the threshold of 12.5, indicating successful reconstruction. However, at $1+\delta > 15$, there is a systematic reduction. Due to the insufficient number of massive halos in the training samples, particularly in the cluster regions, there should be less learning of the halo mass relationship. However, it only occupies 0.01\% grids at the region with $1+\delta > 15$. In comparison, the ratios for thresholds of 12.5 and 13.5 deviate from unity much more for small and large $1+\delta$. Expectedly, as fewer halos trace the underlying mass, UNet is unable to determine how the halo and dark-matter density fields interact. Therefore, we used the halo mass threshold of $12.5$ in subsequent analysis.

\subsection{Power Spectrum}

\begin{figure}
    \centering
	\includegraphics[trim=0cm 3cm 0cm 1.5cm,clip=True,width=1.0\columnwidth]{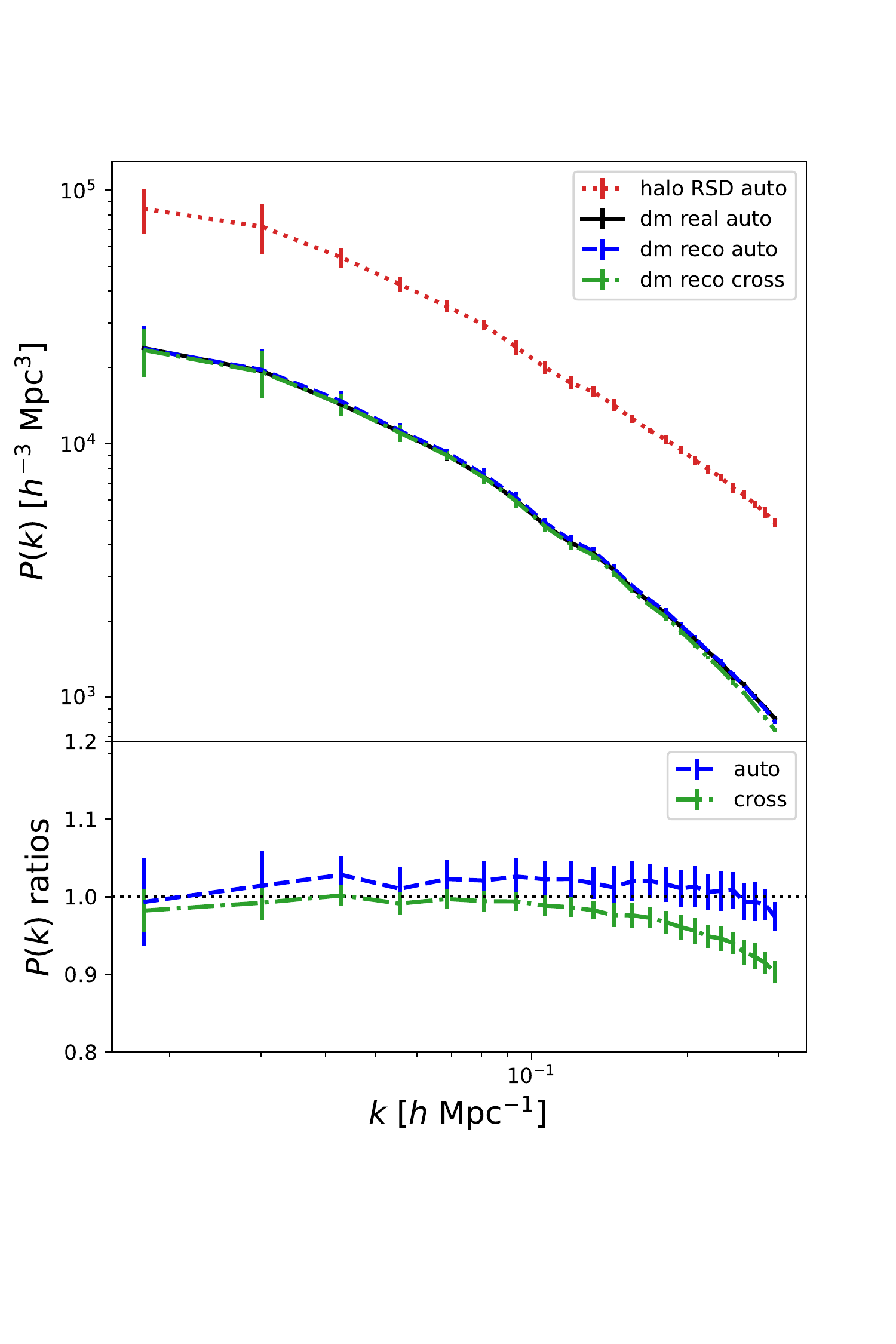}
    \caption{COLA test: Comparison of $P(k)$ (upper panel) and $P(k)$ ratios (lower panel). The black solid and blue dashed lines represent the auto-power spectrum of the true and reconstructed fields, respectively, where the reconstruction is based on the halos with a mass threshold of 12.5. The cross-power spectra of the true and reconstructed fields are shown with the green dashed line. For comparison, the auto-power spectra of the halo density field in the redshift space are shown with the red dashed line. The ratio of the reconstructed and true power spectra is shown in the bottom panel.}
    \label{fig:pk1d_cola}
\end{figure}

The clustering is evaluated by comparing the power spectrum of the reconstructed dark-matter density fields with the truth. In the top panel of Figure \ref{fig:pk1d_cola}, we illustrate the comparisons of the power spectrum. The black solid and blue dashed lines represent the auto-power spectrum of the true and reconstructed density fields respectively, where the reconstruction is based on halos with a mass threshold of 12.5. For comparison, the red dot line shows the auto-power spectra of the redshift-space halo density field. The bottom panel indicates the ratio between the reconstructed and true power spectra, as indicated. The error bars represent the $\pm 1\sigma$ variance among the 10 test samples. As demonstrated, the halo's power spectra is biased to that of dark matter. After reconstruction, the resulting auto-power spectra match the target at the $1\sigma$ level across the range of scales $k < 0.3~h~\mathrm{Mpc}^{-1}$.

However, even if the power spectra are perfectly reproduced, this does not imply that the reconstructions are well performed. In terms of two independent COLA simulations, they have similar power spectra, but their phase space information is significantly different. To check the concordance between the reconstructed and the true density field, we validated the cross-correlation power spectrum, which is referred to as the phase correlation in \citep{2013ApJ...772...63W}. In Figure \ref{fig:pk1d_cola}, the green dashed-dot line represents the cross-power spectra of the true and reconstructed fields, as well as the ratios of the reconstructed cross-power spectra to the true one. It is reassuring that the ratios match the unity perfectly within the $1\sigma$ range at scales $k < 0.1~h~\mathrm{Mpc}^{-1}$. On smaller scales $0.1 < k < 0.3~h~\mathrm{Mpc}^{-1}$, it reveals the value of ratios less than unity. Overall, the reduction of the cross-correlation power spectrum at $k=0.1$ and $0.3~h~\mathrm{Mpc}^{-1}$ are at 1\% and 10\% levels, respectively.

\subsection{Quadrupole}
\label{sec:qua_cola}

\begin{figure*}
    \centering
	\includegraphics[width=1.8\columnwidth]{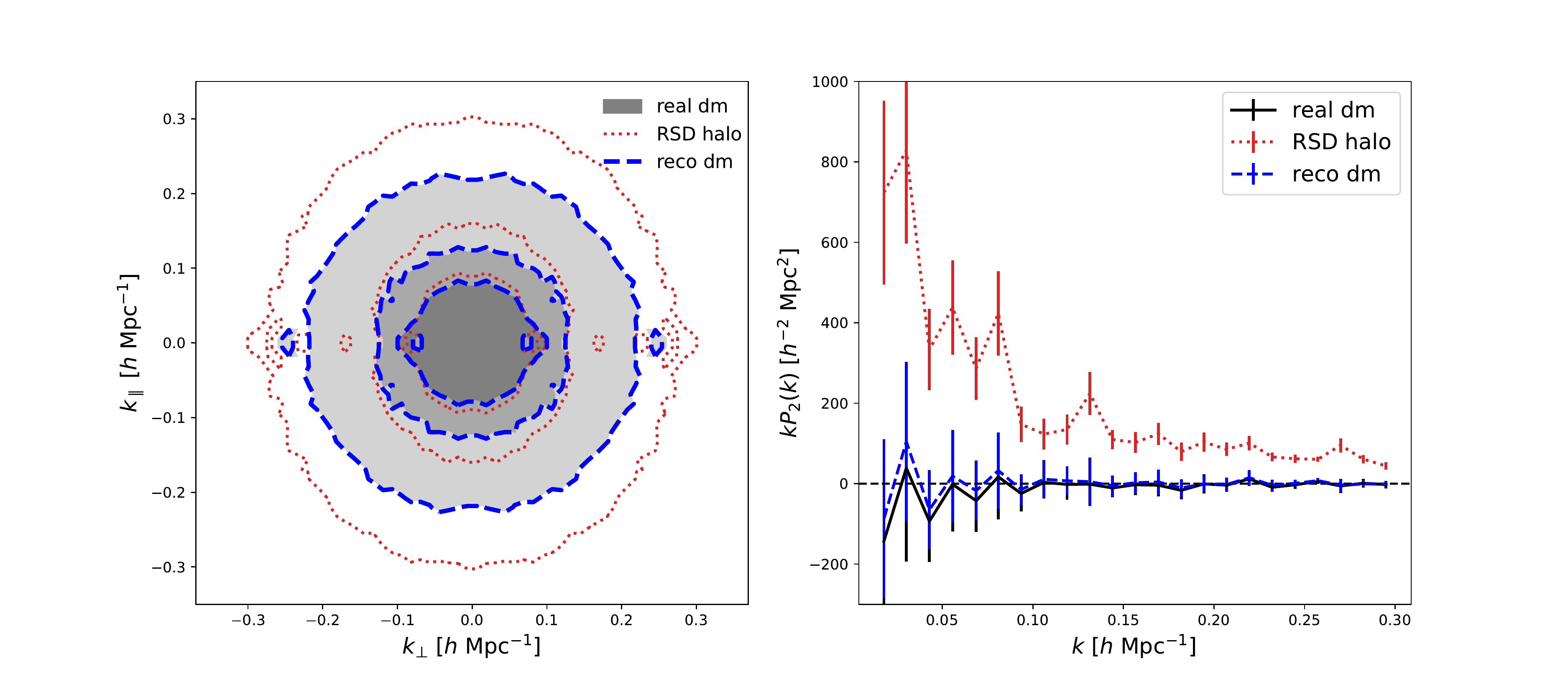}
    \caption{COLA test: (Left panel) Comparison of the two-dimensional (2D) power spectra of the reconstructed and true dark-matter density fields. The gray band and blue dashed contours represent the true and reconstructed fields, respectively. The red dotted contours represent the 2D power spectra of the redshift-space halo with a factor $1/b^2$, where $b$ is the bias factor of halos. Note that the results are averaged for the 10 test samples. (Right panel) Comparison of the quadrupole of the reconstructed (blue dashed) and true (black solid) density fields. The red dotted line represents the results for the redshift-space halos. Note that density reconstruction is based on the halos with a mass threshold of 12.5.}
    \label{fig:pk2d_cola}
    \hspace*{0cm}\includegraphics[trim=0cm 2cm 0cm 2cm,clip=True,width=1.6\columnwidth]{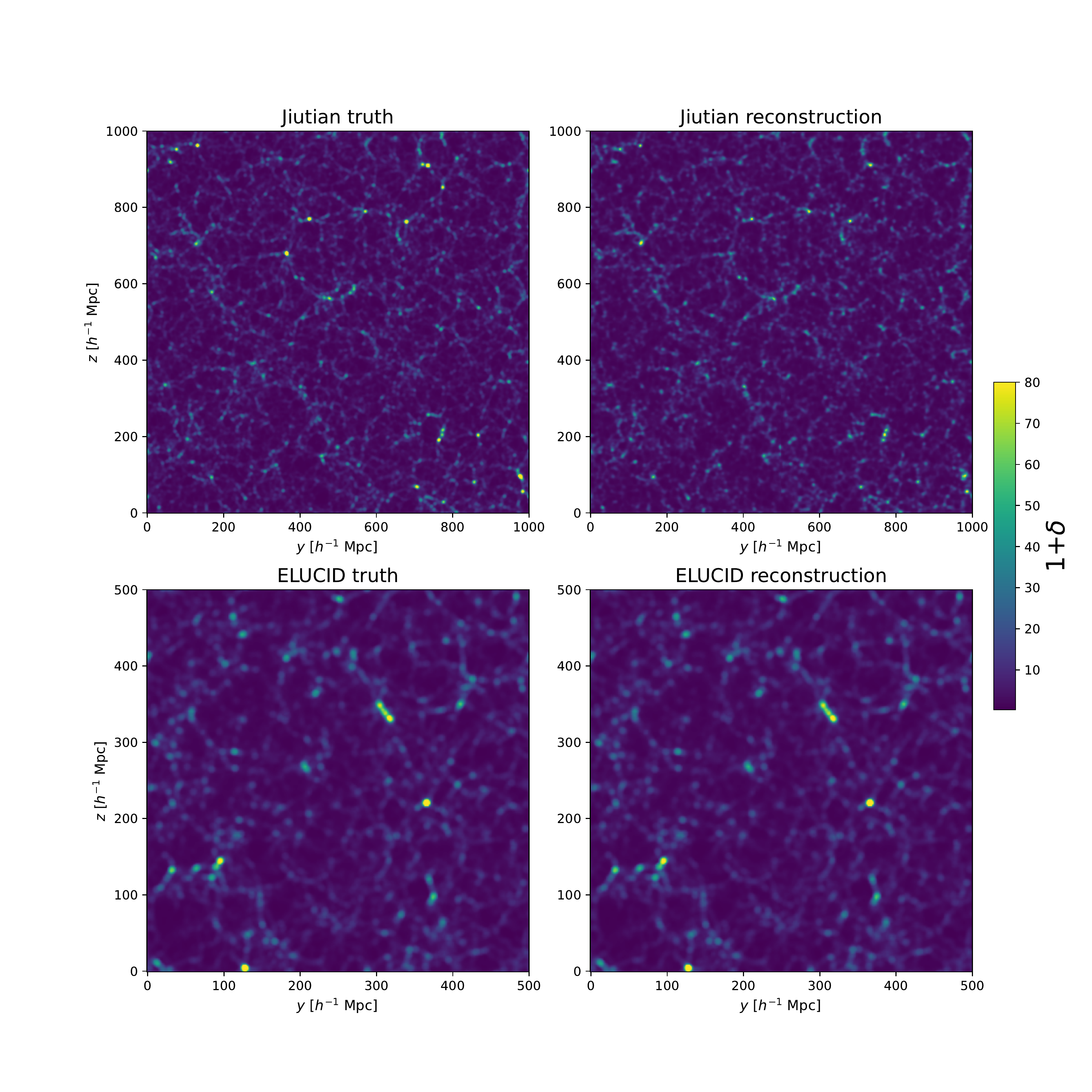}
    \caption{Jiutian and ELUCID test: Projected true density fields (left panels) versus the reconstructed fields (right panels). The upper and lower panels represent Jiutian in a slice of $1000 \times 1000 \times 9.76 \mpch$ and ELUCID in a slice of $500 \times 500 \times 9.76 \mpch$, respectively.}
    \label{fig:slices_jiutian}
\end{figure*}

To assess the accuracy of the correction of the RSDs, we compared the 2D power spectra, $P(k_\bot,k_\parallel)$, of the reconstructed dark-matter density field with the truth, as shown in the left panel of Figure \ref{fig:pk2d_cola}. The gray band and blue dashed contours represent the true and reconstructed field, respectively. The red dot contours represent the redshift-space halo's $P(k_\bot,k_\parallel)$ with a factor of $1/b^2$, where $b$ is the bias factor of halos. It should be noted that the results are an average of the 10 test samples. The redshift-space $P(k_\bot,k_\parallel)$ is clearly anisotropic, with the ``elongated'' feature in the central region, revealing the impact of the Kaiser effect on large scales (small k values). After reconstruction, the resulting $P(k_\bot,k_\parallel)$ is unambiguously more isotropic and perfectly round on large scales. A comparison with the true $P(k_\bot,k_\parallel)$ demonstrates that the correction for the RSDs is generally successful.

The right panel of Figure~\ref{fig:pk2d_cola} compares the quadrupole, $P_2(k)$. The back and blue dashed lines represent the true and reconstructed $P_2(k)$, respectively. The error bars represent the variance among the 10 test samples. The red dot line shows the outcome of the redshift-space halo. In redshift space, the Kaiser effect causes the quadrupole to deviate significantly from zero. In the true field without RSDs, we expect isotropy to result in a quadrupole $P_2(k)=0$. Thus, if the redshift distortion correction is successful, the resulting clustering should have a vanishing quadrupole, and thus $P_2(k)=0$. As illustrated, in redshift space, $P_2(k)$ deviates significantly from zero, whereas in the true field, $P_2(k)$ is close to zero. In the reconstruction, the quadrupole is much closer to zero within the error bars o at various scales, indicating that the reconstruction successfully corrected the majority of RSDs.

Based on the test results, we conclude that our NN with a small sample size is capable of predicting the density field for new data and can accurately correct for RSDs, resulting in a reliable clustering dark-matter density field. It should be noted that additional training sets are still required to improve performance at small scales. To obtain a more accurate reconstruction, reduce the variance of the model prediction by averaging several models saved from the NN.

\begin{figure*}
    \centering
	\includegraphics[trim=4cm 0cm 0cm 0cm,clip=True,width=2.2\columnwidth]{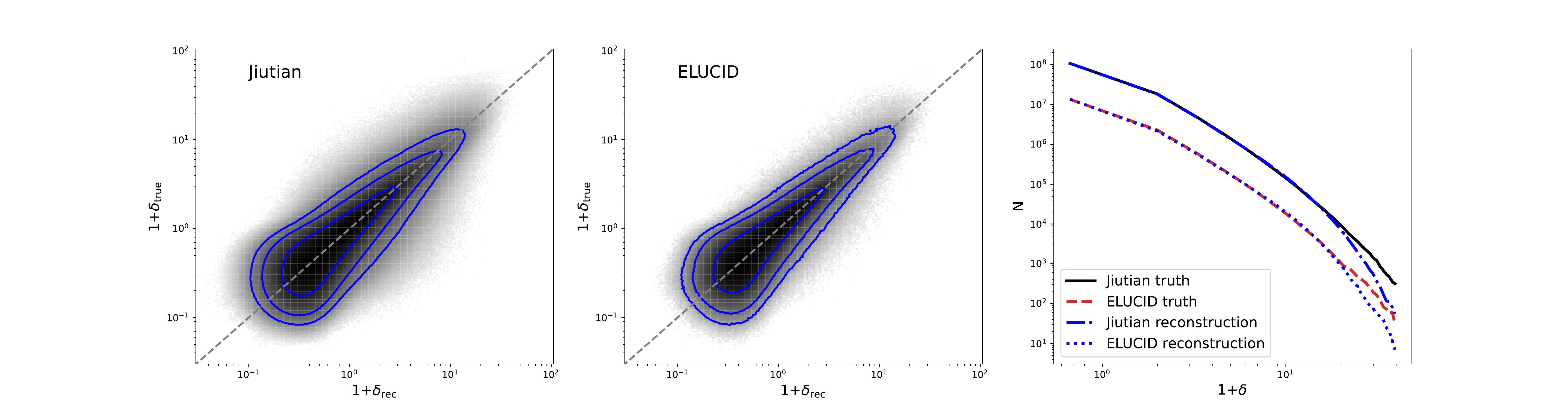}
    \caption{Jiutian and ELUCID test: (Left and middle panels) Density--density relationship exists between the reconstructed density field $1+\delta_\mathrm{rec}$ and the true field $1+\delta_\mathrm{true}$. The three contours cover 67\%, 95\%, and 99\% of the grid cells in the reconstruction volume. The dashed lines represent the perfect relationship. (Right panel) Comparison of the voxel number distributions. The back solid and red dashed lines represent the true density fields obtained in the Jiutian and ELUCID simulations, respectively. The blue dash-dotted and dotted lines represent the reconstruction for Jiutian and ELUCID, respectively.}
    \label{fig:pix_jiutian}
	\includegraphics[trim=0cm 4cm 0cm 1.5cm,clip=True,width=2.0\columnwidth]{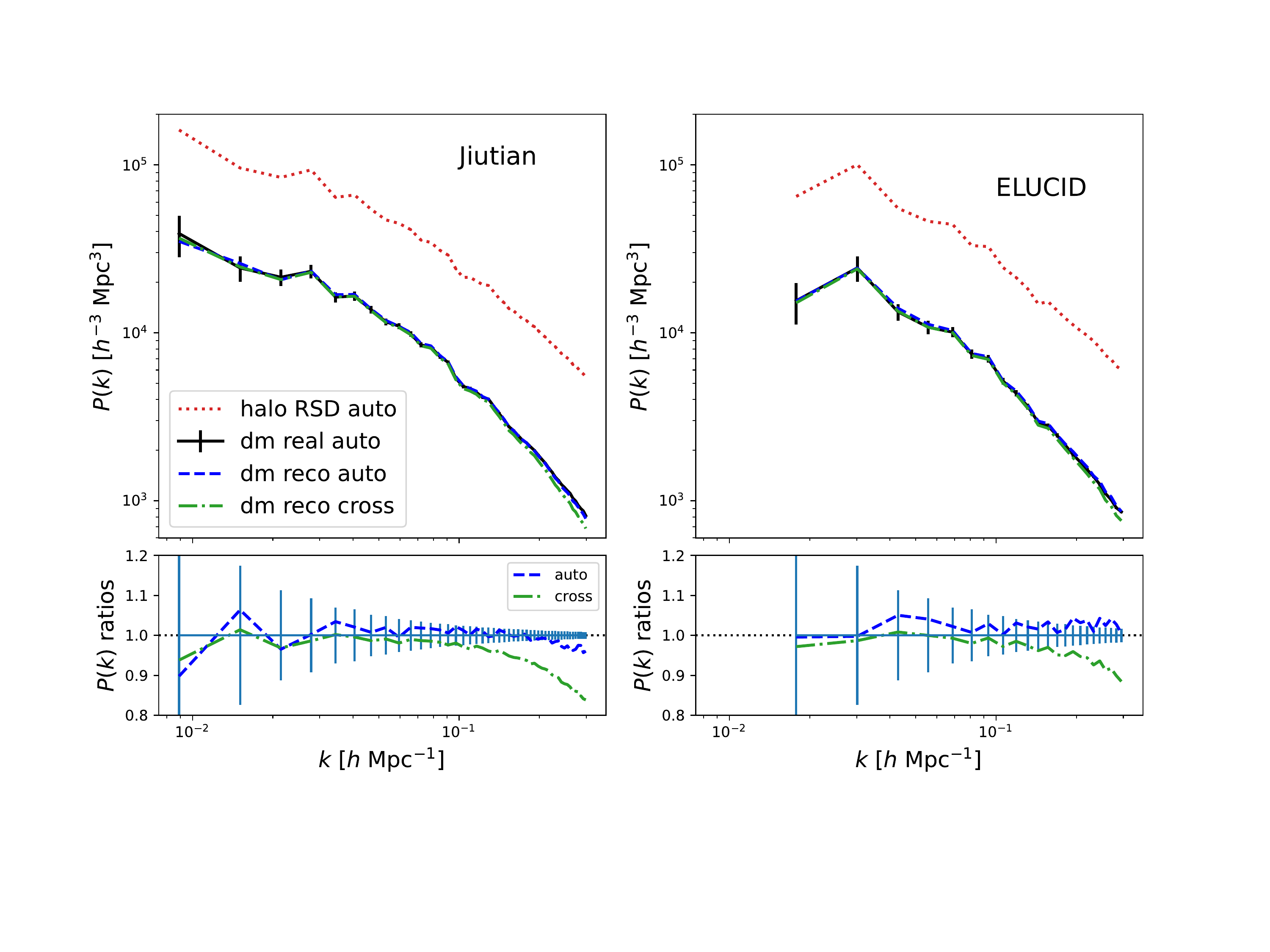}
    \caption{Jiutian and ELUCID test: Identical to Figure~\ref{fig:pk1d_cola} but for Jiutian (left panels) and ELUCID (right panels). The error bars are computed using Equation~\ref{eq:pker}.}
    \label{fig:pk1d_jiutian}
\end{figure*}

\vspace{3cm}

\section{Validation with N-body simulations}
\label{sec:val_jiutian}
In this section, our goals are to (i) test the generality of the COLA-trained UNet model to new sets of samples from the N-body simulation with much larger volume and higher resolution and (ii) investigate the impact of cosmology on the reconstruction accuracy. Therefore, we validated the reconstruction using Jiutian and ELUCID N-body simulation data, as described in Section~\ref{sec:data}, where both halo samples were cut using the same mass threshold of $10^{12.5}\msunh$ as the COLA samples. Unlike the COLA training samples, Jiutian has a much larger volume and higher resolution, whereas ELUCID employs distinct cosmological parameters. To match the box volume of the COLA training sample, we divided Jiutian into eight identically sized boxes before applying the UNet model and then concatenated them back to the box in the original volume.

As a visual comparison, Figure~\ref{fig:slices_jiutian} compares the projected true density fields (left panels) to the reconstructed fields (right panels). The upper and lower panels represent Jiutian in a slice of $1000 \times 1000 \times 9.76 \mpch$ and ELUCID in a slice of $500 \times 500 \times 9.76 \mpch$, respectively. As can be seen, both reconstructed density fields exhibit filamentary structures connecting high-density nodes, similar to that in the original field. Figure~\ref{fig:pix_jiutian} shows the density-density relationships and the comparisons of the density distributions between the reconstructed and true density fields for both Jiutian and ELUCID samples, as indicated. The reconstructed density fields are linearly correlated with the true fields and do not show any significant bias in either case, which is consistent with the result from the COLA test sample results. In the density distribution, the reconstruction correlates favorably with the true field at the appreciable range despite an underestimation at large $\delta$, indicating overall success for the reconstruction of both Jiutian and ELUCID samples. Although reconstructions are systematically underestimated in the range of $1+\delta>20$, it only accounts for 0.02\% of total voxels. The performance of the reconstruction could be attributed to a lack of massive halos in the training samples, which resulted in poor learning for the halo-matter relation at high-density regions, as discussed in Section~\ref{sec:cola_pixcom}.

Figure~\ref{fig:pk1d_jiutian} shows the comparisons of the power spectrum, $P(k)$, for Jiutian (left panels) and ELUCID (right panels). As indicated, different lines are shown for the auto-correlation and cross-correlation $P(k)$ of true and reconstructed dark matter, as well as the auto-correlation $P(k)$ of redshift-space halo density fields. The bottom panels indicate the ratios between the reconstructed and true power spectra. By considering the cosmic variance due to the limited number of Fourier modes for a finite-volume survey, we estimated the error for the power spectrum in a theoretical way \citep{2021PhRvD.104d3528S},
\begin{equation}\label{eq:pker}
    \Delta P(k) = \frac{P(k)}{\sqrt{N_k}}
\end{equation}
%

\begin{figure*}
    \centering
	\includegraphics[trim=0cm 2cm 0cm 3.5cm,clip=True,width=1.5\columnwidth]{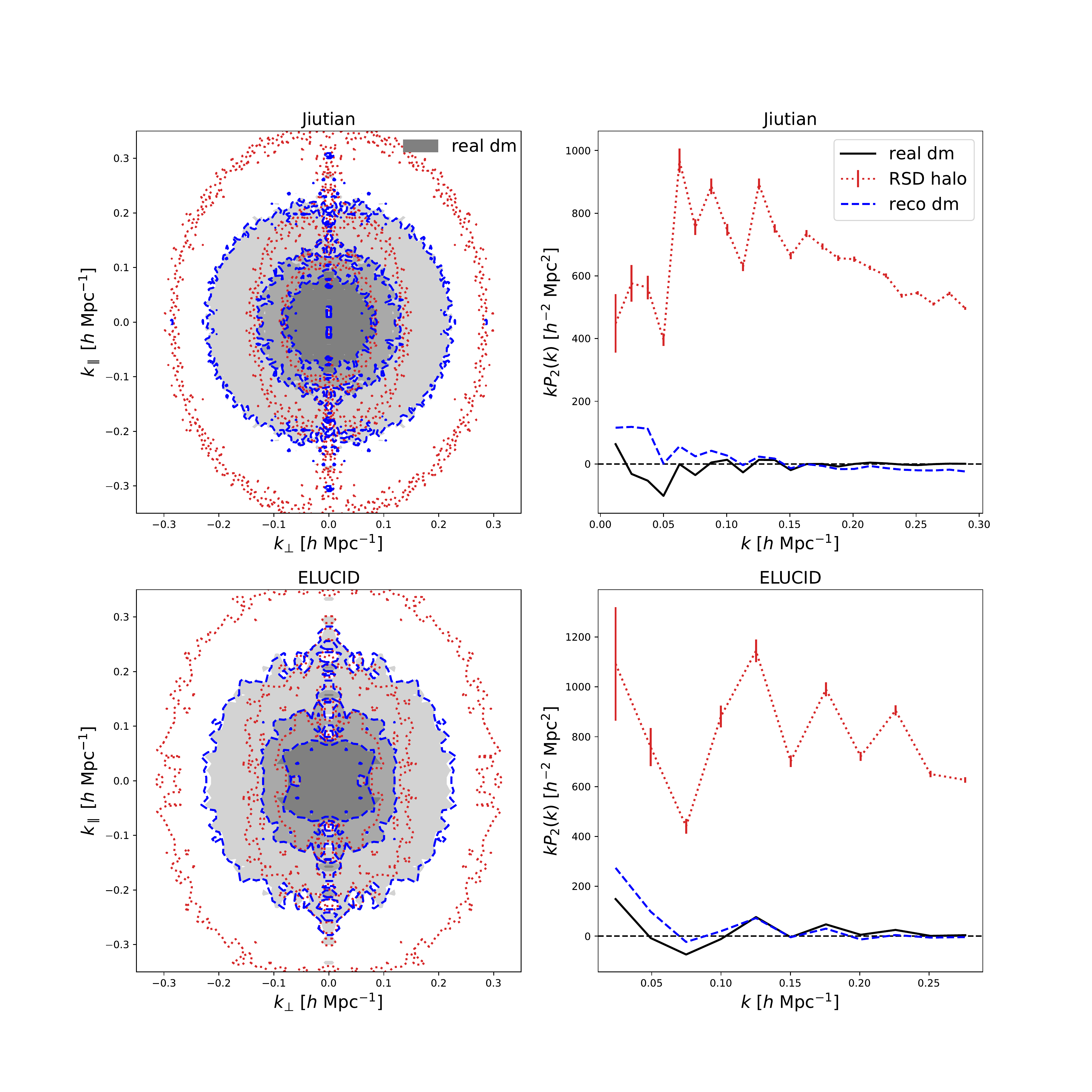}
    \caption{Jiutian and ELUCID test: identical to Figure~\ref{fig:pk2d_cola} but for Jiutian (upper panels) and ELUCID (lower panels). The error bars for the quadrupole of the redshift-space halo density field are computed using Equation~\ref{eq:pker}.}
    \label{fig:pk2d_jiutian}
\end{figure*}
\noindent where $N_k$ denotes the number of independent $k$-modes available per bin. When compared to the halos' $P(k)$, the reconstructed auto-correlation $P(k)$ values match the target truths at the $1\sigma$ level over the range of scales $k < 0.3~h~\mathrm{Mpc}^{-1}$. The cross-correlation $P(k)$ values lie between the auto-power spectrum, and the ratios are close to unity at the $1\sigma$ level on a scale of $k < 0.1~h~\mathrm{Mpc}^{-1}$. The underestimation at scale $k > 0.1~h~\mathrm{Mpc}^{-1}$ is likely due to inaccurate reconstruction in the high-density region due to the limited number of massive halos in current training samples (see also the discussion in Section~\ref{sec:cola_pixcom}). More importantly, the overall reduction of the cross-correlation power spectrum at $k=0.1$ and $0.3 h \mathrm{Mpc^{-1}}$ are similar to COLA testing results, again at 1\% and 10\% levels, respectively.
As a result, we conclude that the reconstructions of the dark-matter power spectrum for the Jiutian and ELUCID simulations are both appreciably precise and consistent with the COLA test results.

To test the correction for the RSD effect, Figure~\ref{fig:pk2d_jiutian} compares the 2D power spectrum $P(k_\bot,k_\parallel)$ (left panels) and the quadrupole $P_2(k)$ (right panels). The upper and lower panels represent Jiutian and ELUCID results, respectively. The black and blue colors are indicated for the true and reconstructed density fields, respectively. For example, the red dashed lines represent the redshift-space halo's $P(k_\bot,k_\parallel)$, which is multiplied by the factor $1/b^2$ (with $b=1.86$ serving as the halo bias parameter at redshift $z=0$). Like the conclusions based on COLA test samples, the reconstructed $P(k_\bot,k_\parallel)$ values are in good agreement with the true $P(k_\bot,k_\parallel)$ and much more isotropic than the redshift-space halos. In the quadrupole, as expected, the redshift-space halo's $P_2(k)$ values deviate significantly from zero, whereas the $P_2(k)$ values for the reconstruction are much closer to zero. When compared to the true $P_2(k)$ values, the reconstructed $P_2(k)$ values have a modest positive signal on large scales, which is due to noise according to the test in Section~\ref{sec:qua_cola}. On smaller scales, the reconstruction shows a slightly negative $P_2(k)$, which could be due to an inaccurate reconstruction in the high-density region. As a result, our UNet-based reconstruction can effectively eliminate RSD effects for both Jiutian and ELUCID samples.

Therefore, we conclude that the UNet model, which was fine-tuned using COLA training samples of low volume and low resolution, is capable of accurately reconstructing the dark-matter density field of the N-body simulation even with larger volume and higher resolution. Meanwhile, the differences in cosmologies between the training (Planck 2018) and test samples (WMAP5) do not affect our reconstruction.

\begin{figure*}
    \centering
	\hspace*{-2cm}\includegraphics[width=2.5\columnwidth]{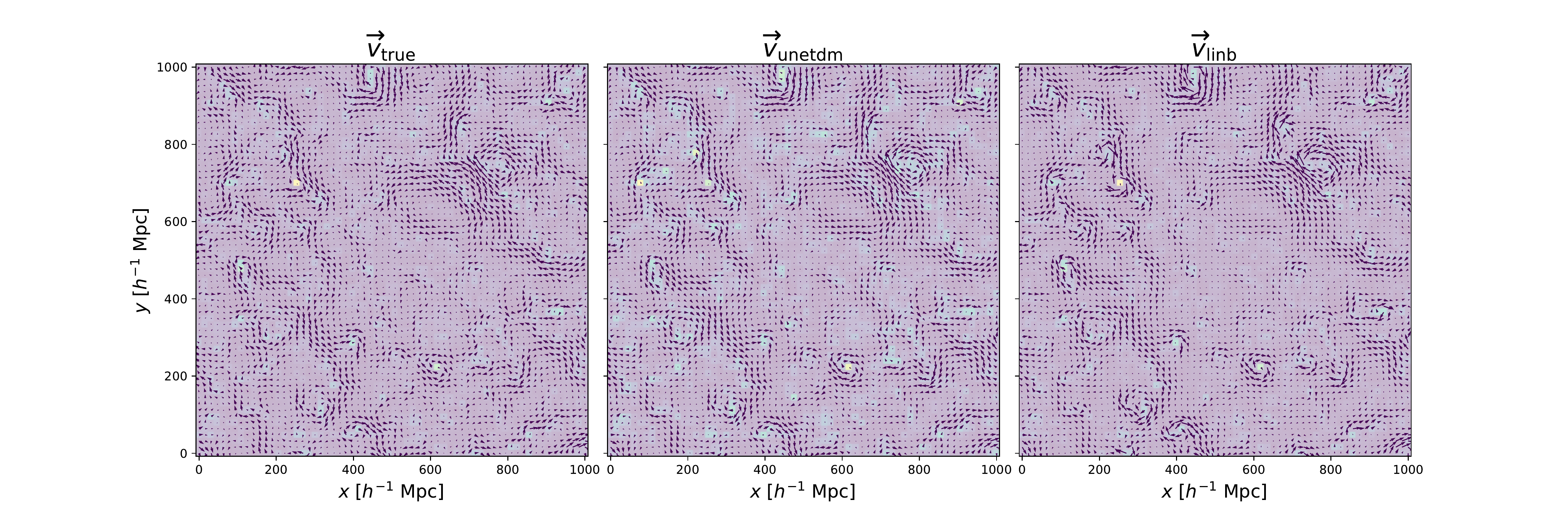}
    \caption{Jiutian test: Velocity and density fields in a slice of $1000 \times 1000 \times 9.76 \mpch$. The left panel shows the true field, $\vec{v}_\mathrm{true}$, obtained from the original simulation. The middle panel shows the reconstructed dark-matter velocity field, $\vec{v}_\mathrm{unetdm}$, which is derived from the UNet-reconstructed dark-matter density field. The right panel shows the reconstructed dark-matter velocity field, $\vec{v}_\mathrm{linb}$, which was derived from the halo density field with a linear bias $b_{\rm h}$. The length of an arrow is proportional to the magnitude of the velocity it represents.}
    \label{fig:vec_distri}
	\includegraphics[width=2.0\columnwidth]{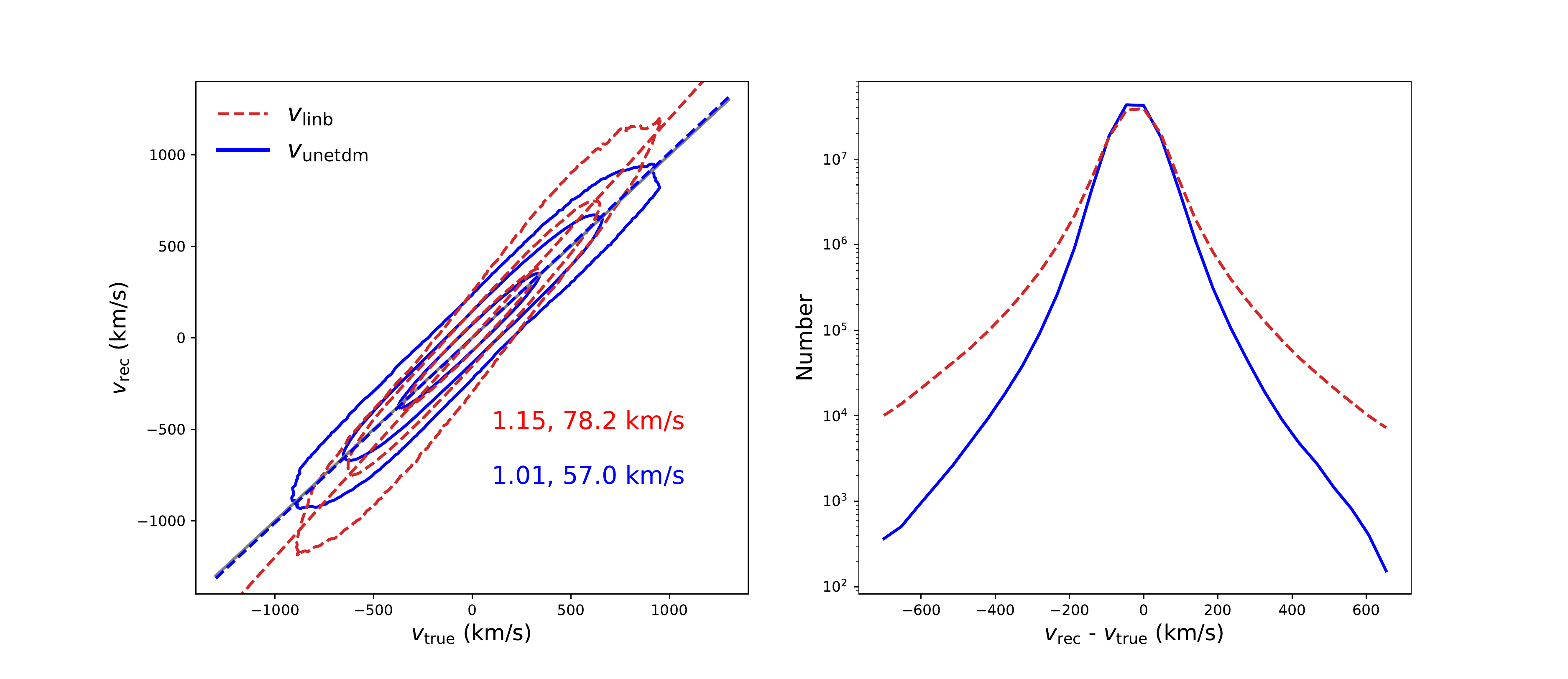}
    \caption{Jiutian test: Comparison of the velocity--velocity relationship (left panel) and the number distribution of the difference $v_\mathrm{rec} - v_\mathrm{true}$ (right panel). The blue solid and red dashed contours represent the reconstructed dark matter velocities $v_\mathrm{unetdm}$ and $v_\mathrm{linb}$, respectively, where the former is recovered from the UNet-reconstructed density field and the latter from the halo density field with a linear bias. Note that all the velocities are along the z-axis. In the left panel, the three contours cover 67\%, 95\%, and 99\% of the grid cells, and the best-fitting relationship of the correlation between reconstruction and simulation is indicated by the corresponding colored dashed lines. The gray line represents the unity slope relationship. The slope of the best fit and the scatter, in terms of the root-mean-square (RMS) in $v_\mathrm{rec} - v_\mathrm{true}$, are indicated in the left panel, with a red dotted line for $v_\mathrm{linb}$ and a blue line for $v_\mathrm{unetdm}$.}
    \label{fig:vec_hist}
\end{figure*}

\section{Testing the reconstruction of the velocity and tidal field}
\label{sec:val_vectid}
We test the effectiveness of the UNet-based reconstructed density field in recovering the velocity field and tidal fields using the Jiutian sample.

\subsection{Reconstructing the velocity field}
The velocity field is computed from the density field reconstructed by the UNet models. In the linear regime, the peculiar velocities are caused by and proportional to the perturbations in the matter distribution. In Fourier space, we have
\begin{equation}\label{eq:vk}
\bm{v}(\bm{k}) = H \, a \, f(\Omega_\rmm) \, \frac{i\bm{k}}{k^2} \, \delta(\bm{k}),
\end{equation}
where $H = \dot{a}/a$ denotes the Hubble parameter, $a$ denotes the scale factor, $f(\Omega_\rmm) = \Omega_\rmm^{0.55}$ \citep{2007APh....28..481L} denotes the growth factor, and $\delta(\bm{k})$ denotes the Fourier transform of the density perturbation field $\delta(\bm{x})$. Hence, for a given cosmology, the linear velocity field can be directly inferred from the density perturbation field, $\delta(\bm{x})$. In the traditional method \citep{2009MNRAS.394..398W}, the peculiar velocity field is reconstructed from the halo density field by substituting $\delta(\bm{k})$ in Eq.~(\ref{eq:vk}) with $\delta_{\rm h}(\bm{k}) / b_{\rm h}$, where $\delta_{\rm h}$ is the dark-matter halo density field and $b_{\rm h}$ is the linear bias parameter for dark-matter halos with mass above a certain mass threshold. In this case, the distribution of halos is considered to be a point set that represents the underlying mass distribution with a local bias. However, the point set of halos may not be a local sampling of the underlying mass density field, as the probability of a halo at a particular location may depend on the presence of other dark-matter particles at or near this location \citep{MBWbook}. In this case, the linear bias method may fail to predict the nonlinear velocity field at small scales. However, although \citep{2009MNRAS.394..398W} proposed a method for predicting the peculiar velocities using the reconstructed density field, it is quite time-consuming because the positions of the redshift-space halos should be corrected and iterated until the convergence is achieved. Herein, we successfully reconstructed the density field by correcting for the RSD effects and directly obtained the velocity field using Eq.~(\ref{eq:vk}) without making any assumptions about the linear bias.

In Figure~\ref{fig:vec_distri}, we show the velocity and density fields in a slice of $1000 \times 1000 \times 9.76 \mpch$. The left panel shows the true field ($\vec{v}_\mathrm{true}$) from the original simulation, while the middle panel shows the reconstructed dark-matter velocity field ($\vec{v}_\mathrm{unetdm}$), which was derived from the UNet-reconstructed density field. For comparison, the right panel shows the reconstructed dark-matter velocity field, $\vec{v}_\mathrm{linb}$, which was derived from the halo density field with the linear bias $b_{\rm h}$. To avoid iterative correction of RSD effects, we computed $\vec{v}_\mathrm{linb}$ using the halo density field in real space rather than the redshift space, whereas $\vec{v}_\mathrm{unetdm}$ was computed using the UNet-based reconstructed density field recovered from the redshift-space halo density field. In the figures, the length of an arrow is proportional to the magnitude of the velocity it represents. When the density is low, both reconstructions perform well. Nonetheless, the velocities $\vec{v}_\mathrm{linb}$ appear to be significantly noisier in high-density regions, such as those around $(x,y) = (220,700)$. This is an example of inaccuracies for grid cells where nonlocal bias effects are significant. Instead, it is clear that the velocities $\vec{v}_\mathrm{unetdm}$ closely follow the true velocities even in the high-density regions.

In Figure~\ref{fig:vec_hist}, we illustrate the velocity--velocity relation (left panel) and the distribution of the velocity difference (right panel) between the reconstruction and truth, $v_\mathrm{rec} - v_\mathrm{true}$. The blue solid and red dashed contours represent the reconstructed dark-matter velocities $v_\mathrm{unetdm}$ and $v_\mathrm{linb}$, respectively, where the former is recovered from the UNet-reconstructed density field and the latter from the halo density field with linear bias. The three contours cover 67\%, 95\%, and 99\% of the grid cells. Here, we compare only the z-component of the velocities because the RSD effects are assumed to be along the z-axis, and the results for all three directions are comparable. We performed a linear regression, and the best-fitting lines are shown in the figure as blue and red dashed lines for both cases. Additionally, the panel shows the slope and scatter values in terms of the RMS in $v_\mathrm{rec} - v_\mathrm{true}$. The velocity $v_\mathrm{unetdm}$ is linearly correlated with the true velocity with a best-fitting slope of 1.01; Meanwhile, for the velocity $v_\mathrm{linb}$, the best-fitting slope of correlation (1.15) deviates significantly from unity. For the distributions of $v_\mathrm{rec} - v_\mathrm{true}$, both reconstructed velocities peak at zero but the scatter ($78.2$ km/s) of $v_\mathrm{halo}-v_\mathrm{true}$ is 37.2\% larger than that ($57$ km/s) of $v_\mathrm{unetdm}-v_\mathrm{true}$. Notably, the velocity $v_\mathrm{linb}$ is reconstructed without including the error induced by the RSD effects. Based on these findings, we conclude that the UNet-reconstructed dark-matter density field can be reliably used to reproduce the cosmic velocity field, and the UNet-based reconstruction outperforms the halo-based reconstruction.

\begin{figure*}
    \centering
	\includegraphics[width=2.0\columnwidth]{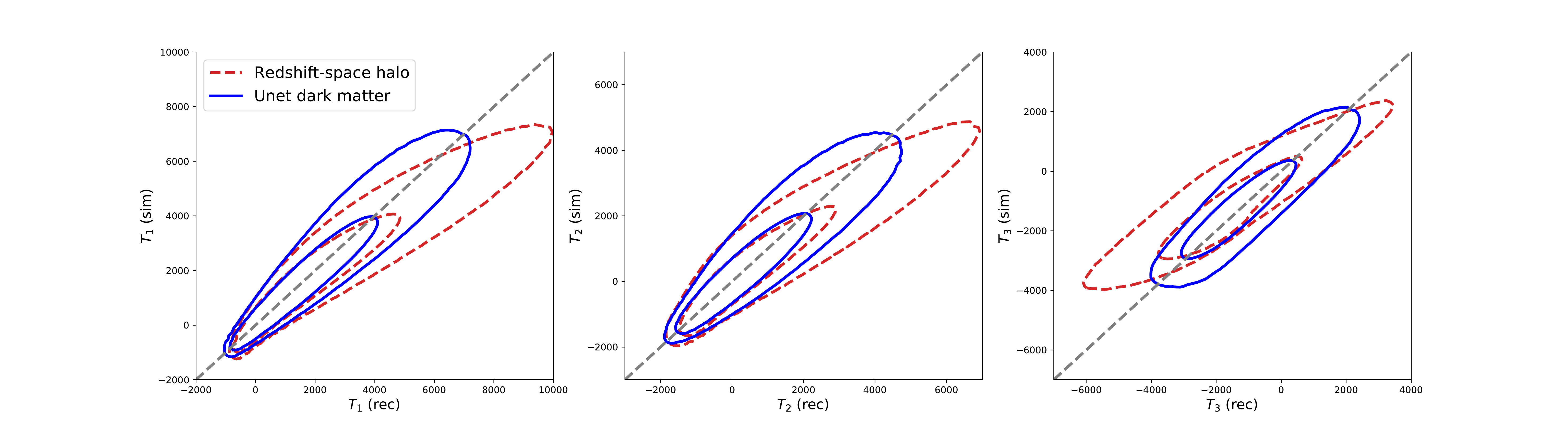}
    \caption{Jiutian test: Eigenvalues of the reconstructed tidal tensor versus those obtained from the true dark-matter density field of the Jiutian simulation. The red and blue contours represent the UNet-reconstructed dark-matter density field and the redshift-space halo density field, respectively. The contours for each result cover between 90\% and 99\% of the grid cells.}
    \label{fig:tidal_relation}
	\includegraphics[width=2.0\columnwidth]{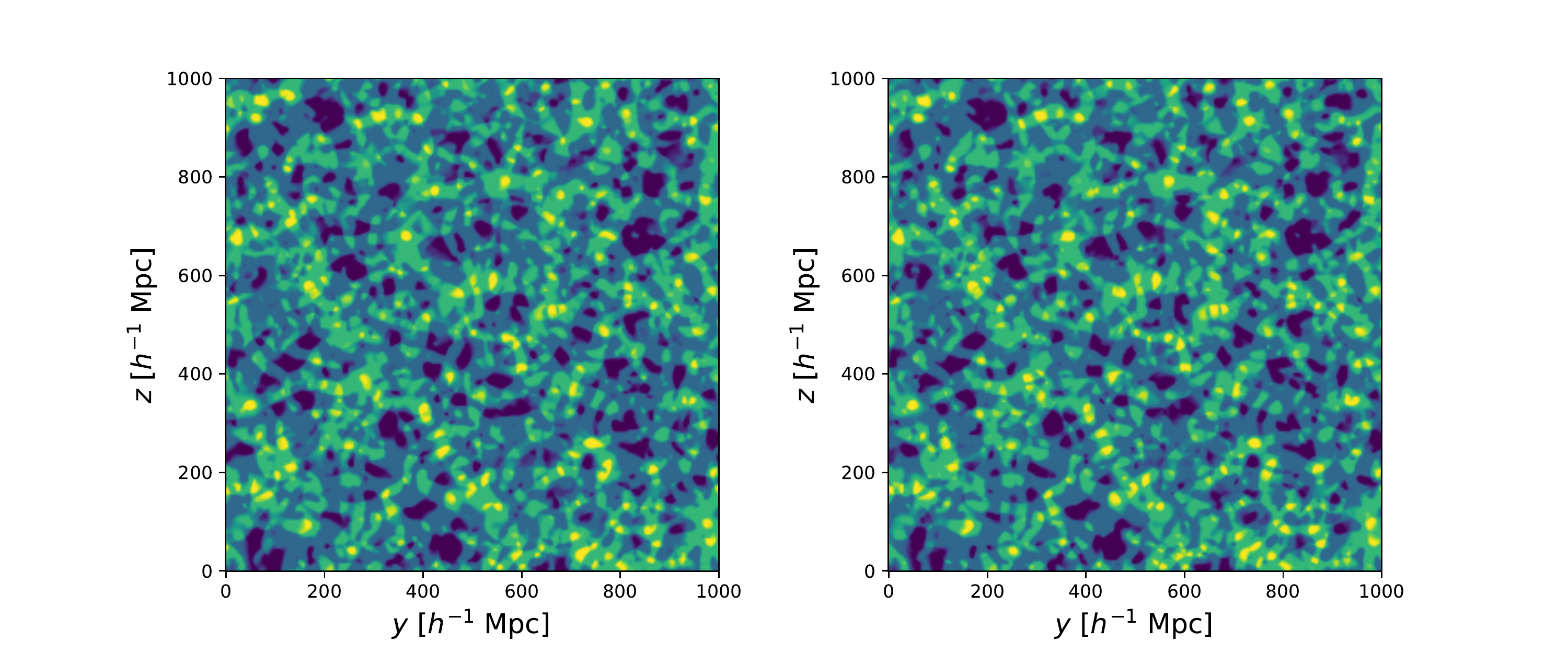}
    \caption{Jiutian test: Comparison of the classification of the large-scale structure between the true dark-matter density field (left panel) and UNet-reconstructed density field (right panel) in a slice of $1000 \times 1000 \times 9.76 \mpch$. The yellow, yellow-green, green, and black regions are grid cells located at structures classified as clusters, filaments, sheets, and voids, respectively.}\label{fig:dmenv}
\end{figure*}
\subsection{Reconstructing tidal field}

We now proceed to the reconstruction of the large-scale tidal field using the tidal tensor, $T_{ij}$, defined as,
\begin{equation}
    T_{ij} = \frac{\uppartial^2\phi}{\uppartial x_i\uppartial x_j},
\end{equation}
where $i$ and $j$ are indices with values of 1, 2, or 3, and $\phi$ is the peculiar gravitational potential calculated from the mass density field using the Poisson equation,
\begin{equation}
    \nabla^2\phi = 4\pi G \bar{\rho}\delta,
\end{equation}\label{eq:poisson}
where $\bar{\rho}$ is the average density of the universe. The tidal tensor is then diagonalized to obtain the eigenvalues $\lambda_1 \geqslant \lambda_2 \geqslant \lambda_3$ of the tidal tensor at each grid point.

Figure.\ref{fig:tidal_relation} compares the eigenvalues of the tidal field reconstructed from the UNet dark-matter density field to the tidal field calculated directly from the true dark-matter density field of the Jiutian simulation. Overall, the eigenvalues of the reconstructed tidal tensor are highly correlated with the true dark-matter density field. For comparison, we also present results using the method from \citep{2009MNRAS.394..398W}, which computes the peculiar gravitational potential from the distribution of dark-matter halos with a linear bias in Equation~\ref{eq:poisson}. It shows clearly that the halo data reconstruction has a significant bias and a slightly larger scatter. This is due to the inaccurate linear bias on a small scale. When we use the UNet-reconstructed dark-matter density field, the bias between the reconstructed and target tidal fields is nearly nonexistent, and the scatter decreases as well.

The eigenvalues of the tidal field can be used to determine the morphologies of LSS in one of the four classes:
\begin{itemize}
    \item cluster: a grid cell with three positive eigenvalues;
    \item filament: a grid cell with one negative and two positive eigenvalues;
    \item sheet: a grid cell with two negative and one positive eigenvalue;
    \item void: a grid cell with three negative eigenvalues.
\end{itemize}

Figure~\ref{fig:dmenv} shows the classification of the LSS in Jiutian simulation, based on the true dark-matter density field (left panel), versus the UNet-constructed density field (right panel). It demonstrates that two of them are quite consistent and well in line with the cold dark matter (CDM) scenario of structure formation. The filamentary structure that connects the yellow grid cells, which are referred to as clusters, is clearly visible in the yellow-green grid cells, whereas the green cells are sheets with more diffused structures enclosing the yellow-green cells. Black areas can also be seen as void cells.

\section{Summary and conclusion}
\label{sec:con}

Herein, we use a deep NN to reconstruct the underlying mass distribution from the distribution of halos in redshift space. Our reconstruction begins with dark-matter halos because galaxy groups can be used to represent the dark halo population and can be well-defined with halo mass from large redshift surveys of galaxies. Using the $256^3$-voxel density-field map as input, we implement a UNet-style NN with seven convolutional layers and seven transpose convolutional layers with skip connections. We trained the UNet-style NN using the COLA simulation, which is an approximation of the N-body simulation. We save the best model at the optimum phase following an in-depth evaluation of the training process.

We first validate the resulting UNet model with 10 COLA testing samples with the same $512^3$ particles and 500 $\mpch$ box size as the training samples. By performing a series of comparison tests, including density-density relation, density distribution, auto-correlation, cross-correlation, quadrupole, and 2D power spectrum, we discovered that the UNet model can reasonably recover the dark-matter density fields and accurately correct for RSDs. Overall, the reconstructed dark matter matches the target truth with only a 1\% and 10\% reduction in the cross-correlation power spectrum at $k = 0.1~h~\mathrm{Mpc}^{-1}$ and $ 0.3~h~\mathrm{Mpc}^{-1}$, respectively.

The resulting UNet model is further validated using the Jiutian simulation, which is a typical N-body simulation with $6144^3$ particles in a 1000 $\mpch$-size box, and the ELUCID N-body simulation, which has a different cosmology of WMAP5. As a result, the dark-matter density field can be recovered consistently at the same accuracy level, and RSD effects can be effectively eliminated for both Jiutian and ELUCID samples. Therefore, the UNet model, which was fine-tuned using COLA small-volume and low-resolution training samples, is capable of accurately reconstructing the dark-matter density field of the N-body simulation even with larger volume and higher resolution. Meanwhile, the differences in cosmologies between the training sample (Planck 2018) and the test samples (WMAP5) do not affect our reconstruction.

Finally, we evaluate the effectiveness of the UNet-based reconstructed density field in recovering the velocity and tidal fields using the Jiutian simulation data. Our findings are as follows.

\begin{itemize}
    \item The reconstructed velocity field from the UNet-reconstructed density field has a 1.01 slope linear correlation to the true velocity field, and the distributions of $v_\mathrm{rec} - v_\mathrm{true}$ peak at zero with an RMS error of $57$ km/s. In comparison, the velocities reconstructed from the halo density field (with a linear halo bias) show a significant bias with a slope of 1.15 and a 37.2\% larger scatter.
    \item The tidal field derived from the UNet-reconstructed density field is unbiased when compared to the true dark-matter tidal field, whereas the tidal field derived from halo data has a significant bias. The eigenvalues of the UNet-based reconstructed tidal field can show recognizable large-scale structures such as clusters, filaments, sheets, and voids, which are consistent with those found in the true dark-matter density field.
\end{itemize}

As a result, the UNet-based model can reliably reproduce the cosmic density field and extends to the high-resolution N-body simulation after being trained with low-resolution COLA simulations. This reduces the computation time required to generate training data and increases the deep-learning method's applicability to massive amounts of data from large galaxy surveys. Furthermore, it is clear that the deep-learning approach outperforms the traditional linear bias model-based approach and is useful in precisely recovering the velocity field and tidal field of dark matter. This paper only serves as a proof-of-concept; actual data applications will be presented in subsequent papers.

\begin{acknowledgments}
This work is supported by the National SKA Program of China (2022SKA0110200 and 2022SKA0110202),  the National Natural Science Foundation of China (Nos.12103037, 11833005, 11890692), 111 project No. B20019, and Shanghai Natural Science Foundation, grant No. 19ZR1466800. We acknowledge the science research grants from the China Manned Space Project (Grant No. CMS-CSST-2021-A02) and the Fundamental Research Funds for the Central Universities (Grant No. XJS221312). This work is supported by High-Performance Computing Platform of Xidian University.
\end{acknowledgments}

\bibliography{sample631}{}
\bibliographystyle{aasjournal}



\end{document}